\documentclass[twocolumn]{aastex631}

\usepackage{natbib}
\usepackage{amsmath}
\usepackage{xcolor}
\usepackage{hyperref}
\usepackage{academicons}

\newcommand{\eor}{$e$--$\omega$--$\rho$\ }

\shorttitle{Accurate and efficient photo-eccentric transit modeling}
\shortauthors{MacDougall, Gilbert, \& Petigura}

\graphicspath{{./}{Figures/}}

\begin{document}

\title{Accurate and efficient photo-eccentric transit modeling}

\correspondingauthor{Mason G. MacDougall}
\email{macdougall@astro.ucla.edu}

\author[0000-0003-2562-9043]{Mason G. MacDougall}
\affiliation{Department of Physics \& Astronomy, University of California Los Angeles, Los Angeles, CA 90095, USA}

\author[0000-0003-0742-1660]{Gregory J. Gilbert}
\affiliation{Department of Physics \& Astronomy, University of California Los Angeles, Los Angeles, CA 90095, USA}

\author[0000-0003-0967-2893]{Erik A. Petigura}
\affiliation{Department of Physics \& Astronomy, University of California Los Angeles, Los Angeles, CA 90095, USA}

\begin{abstract}

A planet's orbital eccentricity is fundamental to understanding the present dynamical state of a system and is a relic of its formation history. There is high scientific value in measuring eccentricities of \textit{Kepler} and TESS planets given the sheer size of these samples and the diversity of their planetary systems. However, \textit{Kepler} and TESS lightcurves typically only permit robust determinations of planet-to-star radius ratio $r$, orbital period $P$, and transit mid-point $t_0$. Three other orbital properties, including impact parameter $b$, eccentricity $e$, and argument of periastron $\omega$, are more challenging to measure because they are all encoded in the lightcurve through subtle effects on a single observable --- the transit duration $T_{14}$. In \citealt{Gilbert22}, we showed that a five-parameter transit description $\{P, t_0, r, b, T_{14}\}$ naturally yields unbiased measurements of $r$ and $b$. Here, we build upon our previous work and introduce an accurate and efficient prescription to measure $e$ and $\omega$. We validate this approach through a suite of injection-and-recovery experiments. Our method agrees with previous approaches that use a seven-parameter transit description $\{P, t_0, r, b, \rho_\star, e, \omega\}$ which explicitly fits the eccentricity vector and mean stellar density. The five-parameter method is simpler than the seven-parameter method and is ``future-proof'' in that posterior samples can be quickly reweighted (via importance sampling) to accommodate updated priors and updated stellar properties. This method thus circumvents the need for an expensive reanalysis of the raw photometry, offering a streamlined path toward large-scale population analyses of eccentricity from transit surveys.

\end{abstract}


\section{Introduction}
\label{sec:intro}

Out of more than 5,300 confirmed planets to date, $\sim$75\% were discovered via the transit method. These discoveries have paved the way for keystone scientific advancements in our understanding of planet formation, evolution, and demographics. To ensure the reliability of inferences based on the transiting planet population, we must also ensure that characterizations of individual transiting planets are  consistently and accurately derived. Previously, uncertainties on stellar parameters significantly limited the achievable precision of planet properties (e.g. $\sigma(R_\star) \sim 27\%$ and $\sigma(\rho_\star) \sim 51\%$; \citealt{Thompson2018}). Now, in the era of \textit{Gaia} (\citealt{GaiaDR2}) and high-precision stellar characterizations (e.g. $\sigma(R_\star) \lesssim 2\%$ and $\sigma(\rho_\star) \lesssim 10\%$), the determination of key planet properties is limited by light curve modeling \citep[see, e.g.][]{Petigura20}.

A variety of methods exist for modeling transit signals, including various parameterizations (e.g. \citealt{SeagerMallenOrnelas2003}; \citealt{Carter2008}; \citealt{Dawson12}; \citealt{Eastman2013}; \citealt{Thompson2018}; \citealt{Gilbert22}) and sampling techniques (e.g. \citealt{Feroz08}; \citealt{emcee13}; \citealt{exoplanet:2021}; \citealt{dynesty}; \citealt{Gilbert2022}). Differences in posterior inference which arise from adopting a particular model parameterization and sampling method are often assumed to be insignificant relative to other sources of uncertainty. However, if one wishes to achieve percent-level precision on all quantities, one must also carefully consider the strengths and weaknesses of competing model/sampler implementations (see, e.g., \citealt{Gilbert2022}, \citealt{Gilbert22}). Although substantial effort has been put into vetting methods for transit signal \textit{detection} (see, e.g., \citealt{Christiansen15}), far less effort has been devoted to validating subsequent methods for transit signal \textit{modeling}. A key aim of this work - which builds directly upon our previous work in \citealt{Gilbert22}, hereafter G22 - is to place the transit modeling problem on the same secure foundation as the transit detection problem. Our primary focus here is on the effects of model parameterization, with a secondary focus on the role of the sampler.

A popular and straightforward method for transit model parameterization is to use a seven-parameter basis which includes orbital period $P$, transit epoch $t_0$, planet-to-star radius ratio $r$, impact parameter $b$, eccentricity $e$, argument of periastron $\omega$, and either stellar density $\rho_{\star}$ or scaled orbital separation $a/R_{\star}$, these latter two parameters being related via Kepler's third law\footnote{In practice, other parameters related to the stellar limb darkening (i.e. quadratic limb darkening coefficients {$u_1$, $u_2$}) and to the properties of the photometry (i.e. flux zero-point $F_0$ and photometric noise $\sigma_F$) are usually also needed, but these complicating details are not the focus of this paper.} (see, e.g. \citealt{Eastman2013}). This eccentricity-explicit basis $\{P, t_0, r, b, e, \omega, \rho_{\star}\}$, \eor hereafter, benefits from being fully characterized by properties of the star, planet, and planetary orbit. However, real-world photometric transit lightcurves typically only include enough information to constrain four or five out of the seven parameters. More precisely, in most real-world cases the signal-to-noise $S/N$ of observations is low enough that one cannot precisely measure the duration and curvature of ingress/egress nor can one detect any transit asymmetry (see \citealt{Barnes07}). Without resolved ingress/egress or transit asymmetry, the problem remains unconstrained, with $b$, $e$, $\omega$, and $\rho_{\star}$ each imprinting themselves on the lightcurve indirectly via the transit duration $T_{14}$ ($r$ imprints itself via the transit depth; $P$ and $t_0$ via the ephemeris). More explicitly, for a given $\rho_{\star}$, $b$ influences the transit chord length, $e$ and $\omega$ influence the speed of the planet during transit, and the ratio of transit chord length to orbital speed produces $T_{14}$.

An alternative approach that improves upon these limitations of the seven parameter method is to model the lightcurve assuming a circular orbit, $e = 0$, regardless of what the true underlying eccentricity might be (see, e.g., \citealt{SeagerMallenOrnelas2003}; \citealt{Dawson12}). This shortcut reduces the total number of model parameters by two with a trade-off that the transit is now explicitly assumed to by symmetric. Fortunately, for virtually all \textit{Kepler} and TESS class photometry, this assumption does not introduce measurable biases into the analysis. In G22, we explored the effectivity of two different five-parameter bases: $\{P, t_0, r, b, T_{14}\}$ vs $\{P, t_0, r, b, \tilde{\rho}\}$, where $\tilde{\rho}$ is the stellar pseudo-density, i.e. the stellar density inferred from the transit photometry under the (probably false) assumption of a circular orbit. We found that the two bases are equivalent when an appropriate Jacobian transformation is properly applied, but that the latter basis introduces complex, non-intuitive covariances between $b$ and $\tilde{\rho}$. These covariances artificially disfavor $b \gtrsim 0.7$, which propagates through to other parameters, shifting $e$ toward higher values and $r$ toward lower ones. Historically, the use of parameter bases which include $\tilde{\rho}$ has resulted in biased inference, and we consequently recommend avoiding the use of $\tilde{\rho}$ altogether. For the remainder of this work, we therefore do not consider any parameterizations which include $\tilde{\rho}$.

Our preferred model parameterization $\{P, t_0, r, b, T_{14}\}$, hereafter the $T_{14}$ basis, benefits from being intuitive and close to quantities which can be directly measured from the transit photometry, which minimizes the risk of introducing unintended bias. In G22, we demonstrated that this parameterization yields unbiased posteriors on both $b$ and $T_{14}$. In this work, we build on G22 to develop a post-hoc importance sampling routine that enables indirect recovery of $e$ and $\omega$ from direct measurements of $T_{14}$ and an independent external constraint on $\rho_{\star}$ (e.g. from asteroseismology or spectroscopy). To validate our methods, perform injection-and-recovery tests using simulated transit photometry over a grid of transit parameters and compare the performance of our proposed $T_{14}$ + importance sampling approach to the performance of the standard \eor modeling basis. We find that the two methods yield equivalent posterior inferences on $b$, $e$, and $\omega$, with significant improvements to speed and efficiency when using our new approach. Another major advantage of our proposed technique is that it is ``future-proof'' in that it allows us to update estimates of $e$ and $\omega$ as stellar characterization is inevitably updated in the future (e.g. from new \textit{Gaia} data releases) \textit{without requiring a computationally expensive re-run of the transit fits}. In comparison, the usual seven-parameter \eor basis ``bakes in'' a particular value of $\rho_{\star}$ at the time of transit modeling.

We lay out our methodology for lightcurve synthesis and transit injection-and-recovery in \S\ref{sec:procedure}. We then highlight the procedural differences between the \eor method (\S\ref{sec:eor}) and our $T_{14}$ method (\S\ref{sec:umb}). In \S\ref{sec:results}, we analyze the results of our injection-and-recovery tests and compare the performances of the two parameterizations. We provide a summary of our conclusions in \S\ref{sec:conclusion}.

\section{Synthetic Lightcurve Construction}
\label{sec:procedure}

\begin{figure*}[ht]
\centering
\includegraphics[width=0.99\textwidth]{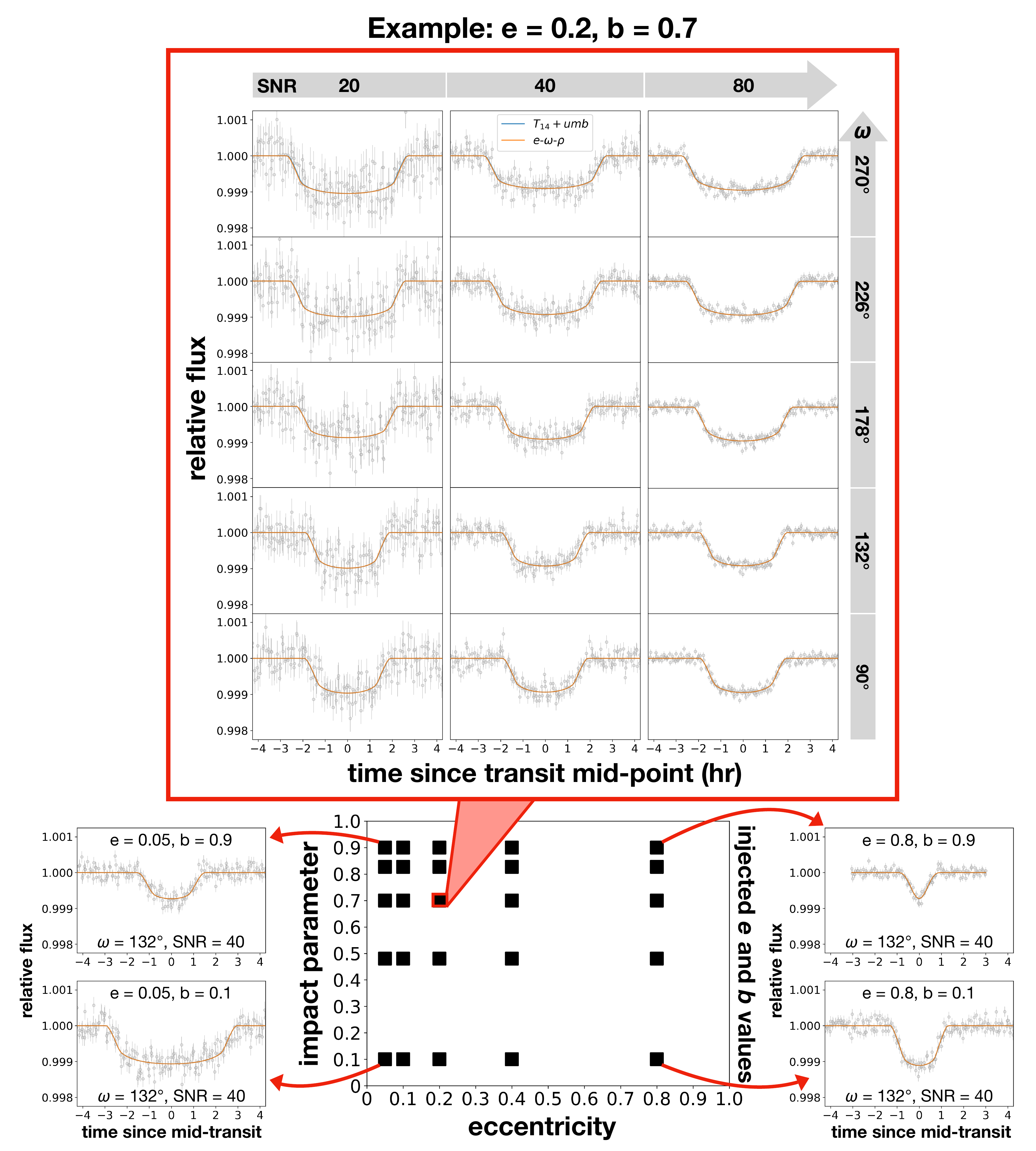}
\caption{Diagram showing (bottom middle) our grid of injected $e$ and $b$ values, along with (top) a gallery of phase-folded lightcurves for all combination of SNR and injected $\omega$ values (at $e = 0.2$ and $b = 0.7$, as an example). The four panels to the left and right of the bottom grid show demonstrative examples of phase-folded lightcurves at the different extremes of our injected parameter grid (e.g. $b = $ 0.1 or 0.9 and $e = $ 0.05 or 0.8), all shown with $\omega = 132^{\rm \circ}$ and SNR = 40. Each lightcurve is shown with the median final transit fits from both the \eor (orange) and $T_{14}+umb$ (blue) modeling methods, which overlap almost entirely.}
\label{fig:lightcurves}
\end{figure*}

Our objective is to compare the performance of the physical \eor parameter basis to the simpler $T_{14}$ basis. We aim to demonstrate whether or not these methods return equivalent and accurate posterior results and determine their relative efficiencies. To achieve these objectives, we perform a suite of injection-and-recovery tests over a grid of parameters which spans a wide range of values of eccentricity $e$, argument of periastron $\omega$, inclination (parameterized as impact parameter $b$), and signal-to-noise (see Figure \ref{fig:lightcurves}). Injection-and-recovery is a standard tool used to evaluate transit signal detection methods (see, e.g., \citealt{Christiansen15}), but it has not been applied to transit model validation on nearly the same scale. Here, we construct a set of synthetic lightcurves, then we proceed to use two distinct transit modeling methods to recover the injected transit properties and compare the relative model performances. 

For all injection-recovery tests in this work, we inject the transit signal of a sub-Neptune-size planet orbiting a Sun-like star with an orbital period close to the average among \textit{Kepler} planets. We synthesize 10 transits per lightcurve with a photometric zero-point flux of $\mu_{\rm flux} = 0$ and a fixed photometric noise $\sigma_{\rm flux}$, consistent with raw photometry that has been accurately prewhitened. We calculate the duration of each injected transit signal $T_{\rm 14}$ according to the following equation from \cite{Winn2010}:

\begin{equation}
\label{eq:duration}
T_{\rm 14} = \frac{P}{\pi}\sin^{-1}\left(\sqrt{\frac{\left(1+r\right)^2-b^2}{\left(\frac{G P^2 \rho_\star}{3 \pi}\right)^{2/3}-b^2}}\right)\frac{\sqrt{1-e^2}}{1+e\sin{\omega}}.
\end{equation}

We construct our synthetic lightcurves at three different signal-to-noise ratio (SNR) levels: SNR $\sim [20, 40, 80]$. We show example lightcurves for each SNR level in Figure \ref{fig:lightcurves}. At an SNR of 20, the injected signal has a slightly lower significance compared to the median \textit{Kepler} planet signal. At the higher SNR levels of 40 and 80, we seek to identify any differences that emerge between our two models as the transit ingress and egress become more distinct from the photometric noise, making $b$ measurements more precise. From the selected SNR and other injected lightcurve properties, we generate Gaussian white noise per lightcurve centered on $\sigma_{\rm flux}$, which we calculate according to:

\begin{equation}
\label{eq:noise}
\sigma_{\rm flux} = \sqrt{\frac{T_{\rm 14, true} N_{\rm transits}}{t_{\rm exp}}} \frac{r_{\rm true}^2}{\textrm{SNR}},
\end{equation}
where $N_{\rm transits}$ is the number of injected transits and $t_{\rm exp}$ is the simulated exposure time of our synthetic lightcurve. The random seed used to generate the synthetic white noise is unique to each injection-recovery test.

We also assign a unique set of transit parameter values \{$b$, $e$, $\omega$\} for each injection-recovery test, where each of these inputs is drawn from a grid of discrete parameter values (see Figure \ref{fig:lightcurves}). We specifically choose a parameter grid that emphasizes the region of parameter space where the $e-\omega-b$ degeneracy is strongest (see, e.g., \citealt{VE15}) since this is where the two parameterizations are more likely to yield differing results. As a result, our injected planet signals do not exactly mirror the distribution of \textit{Kepler} planets, but they do include a broad range of realistic planet characteristics.

Since the transit shape is more sensitive to small changes in $b$ at high values, we select injected values of $b$ with tighter spacing towards higher values, spanning the non-grazing parameter space. We construct an array of $b$ values that are evenly spaced on a reversed log scale: $b \sim [0.1, 0.48, 0.7, 0.83, 0.9]$. We also prefer to use $e$ values that span the range of eccentricities with tighter spacing towards low-to-moderate values, since these are more common. We select an array of possible $e$ values which are evenly spaced on a log scale: $e \sim [0.05, 0.1, 0.2, 0.4, 0.8]$. Additionally, the $\omega$ values that we draw upon for our grid of injected parameters are intentionally selected to include the inflection points of periastron ($\pi/2$ or 90$^{\circ}$) and apastron ($3\pi/2$ or 270$^{\circ}$) along with three roughly evenly spaced values in between: $\omega \sim [90^{\circ}, 132^{\circ}, 178^{\circ}, 226^{\circ}, 270^{\circ}]$. 
\begin{deluxetable}{lcl}
\tabletypesize{\scriptsize}
\tablecaption{Transit Model Parameters and Priors\label{tab:priorstab}}
\tablehead{\colhead{Parameter} & \colhead{Input Value(s)} & \colhead{Prior}}
\startdata
$P$ (d) & 26.1 & fixed \\
$t_0$ (d) & 1.0 & $t_0 \sim$ N(1.0, 0.1) \\
$r$ & 0.03 & log$r \sim$ U(-9, 0) \\
$b$ & [0.1, 0.48, 0.7, 0.83, 0.9] & $b \sim$ U(0, $1+r$) \\
$u_{\rm 1}, u_{\rm 2}$ & $\{0.4, 0.25\}$ & fixed \\
$\mu_{\rm flux}$ & 0 & fixed \\
$\sigma_{\rm flux}$ & derived & fixed per lightcurve \\
$T_{\rm 14}$ (d) & derived & log$T_{\rm 14} \sim$ U(-9, 0) \\
$e$ & [0.05, 0.1, 0.2, 0.4, 0.8] & $e \sim$ U(0, 0.92) \\
$\omega$ ($^{\circ}$) & [90, 132, 178, 226, 270] & $\omega \sim$ U(-90, 270) \\
$\rho_{\rm \star}$ ($g/cm^3$) & 1.41 (e.g. $\rho_{\odot}$) & $\rho_{\rm \star} \sim$ N(1.41, 0.141) \\
\enddata
\tablecomments{All parameters used in the models discussed throughout this analysis, along with their units, input values, and associated priors (if applicable). Priors include normal (N) and uniform (U) distributions. $b$, $e$, and $\omega$ each have five input value options that form a grid of possible injected transit signal properties. We note that the priors on $e$ and $\omega$ can also be represented via the transform $\{e, \omega\} \rightarrow \{\sqrt{e}\sin{\omega}, \sqrt{e}\cos{\omega}\} \sim$ Disk($\sqrt{0.92}$).}
\vspace{-10mm}
\end{deluxetable}

\begin{figure*}[ht]
\centering
\includegraphics[width=0.99\textwidth]{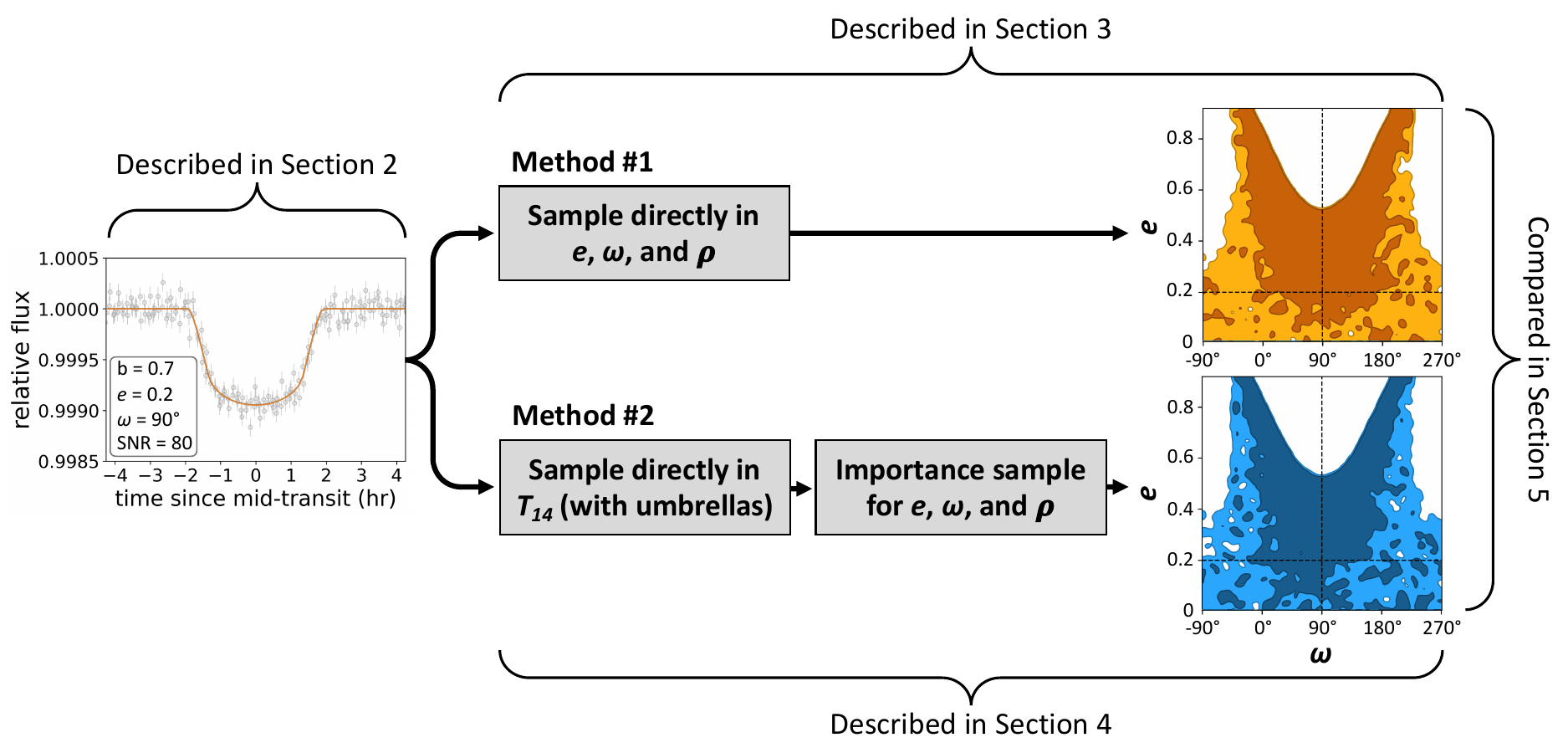}
\caption{Diagram demonstrating an overview of our modeling procedure, from an input synthetic lightcurve to output $e$ and $\omega$ constraints via both the \eor (orange) and $T_{14}+umb$ (blue) modeling methods.}
\label{fig:overview}
\end{figure*}

We construct a set of 375 unique transit lightcurves from all combinations of \{$b$, $e$, $\omega$, SNR\} using the \texttt{batman} transit modeling package (\citealt{batman}). We synthesize these injected lightcurve models with an oversampling rate of 11 and $t_{\rm exp} = 30$ minutes, similar to real \textit{Kepler} photometry. These lightcurves serve as inputs to the two modeling methods that we are comparing, described below, in order to demonstrate similarities and differences in model performance across a range of potential transit signals (see Figure \ref{fig:overview} for an overview).

\section{Method \#1: Direct sampling in $\lowercase{e}$--$\omega$--$\rho$}
\label{sec:eor}

We first model our synthetic transit lightcurves using the \eor model, which serves as our baseline model and standard reference when evaluating the performance of our proposed $T_{14}+umb$ model. This physically-motivated transit model is parameterized by $\{P, t_0, r, b, \rho_\star, e, \omega\}$, along with quadratic limb darkening parameters $\{u_1, u_2\}$. Since we simulate lightcurves with white noise, we fix $\mu_{\rm flux}$ and $\sigma_{\rm flux}$ which would otherwise be directly sampled parameters when modeling real transit photometry.

We construct the \eor model using uninformative priors that are of standard use in transit fitting literature or drawn directly from \textit{G22}, summarized in Table \ref{tab:priorstab}. We apply a normal prior on $\rho_\star$ which assumes that the stellar density is known with $10\%$ uncertainty through independent measurements. To mitigate boundary issues that can occur when sampling $e$ and $\omega$ directly, we use a common redefinition of these parameters $\{\sqrt{e}\sin{\omega}, \sqrt{e}\cos{\omega}\}$ (see, e.g. \citealt{Eastman2013}), with implicit uniform priors on both $e$ and $\omega$. These priors do not account for transit probability or other astrophysically motivated considerations (see \citealt{Barnes07}). 

We implement this model using \texttt{exoplanet} \citep{exoplanet:2021}, with sampling performed by the NUTS algorithm via \texttt{PyMC3} \citep{pymc3:2016}. We use 3,000 tuning steps with an additional 4,000 sampler draws to ensure that the sampler converges with an effective sample size $N_{\rm eff} \approx$ 10$^3$. We also set a high target acceptance fraction of 0.99 to encourage the sampler to adequately explore complex topologies in the posterior parameter space, such as the $b-r$ and $e-\omega$ degeneracies. We follow the standard practice of oversampling the light curve model in order to mitigate binning artifacts (see, e.g., \citealt{Kipping10}), using an oversampling factor of 11. We fit our transit models via two sampler chains across two CPU cores per injection-recovery test.

From initial experimentation, we found that sampler limitations exist which restrict the valid parameter space of eccentricity modeling when applying the \eor parameterization via NUTS sampling with \texttt{exoplanet}. When sampling $e \gtrsim 0.92$, this implementation of the \eor model can have convergence issues due to the high curvature of the posterior parameter space being traversed. This also roughly corresponds with the upper eccentricity limit where we expect transit duration approximations to begin breaking down (see, e.g. \citealt{Kipping2014-asterodensity}). Given that only 5 known planets have $e > 0.9$ and only one of these was discovered via transit modeling, we choose to restrict our eccentricity sampling to $e < 0.92$ for all modeling approaches considered in this work. By doing so, we avoid conflating our primary interest -- differences in modeling methods -- with rare edge cases that are beyond the scope of this work.

\section{Method \#2: Direct sampling in $T_{14}$, then importance sampling in $\lowercase{e}$--$\omega$--$\rho$}
\label{sec:umb}

Our alternative transit modeling approach, the $T_{14}+umb$ model, has a parameter basis that includes the observable transit duration $T_{14}$ as an explicit parameter. This parameterization avoids explicitly sampling the complex degeneracies introduced by $e$ and $\omega$, allowing us to instead measure these parameters post-hoc via importance sampling (see \ref{sec:estimation}). We couple this duration-based parameterization with umbrella sampling (see \citealt{Gilbert2022}) to ensure that our model accurately samples the complicated topology of the high-$b$ ``grazing'' parameter space. Based on the arguments made in both \textit{G22} and \cite{Gilbert2022}, we expect that our $T_{14}+umb$ approach should achieve results that are consistent with those from the \eor model with a potential boost in efficiency.

\subsection{Transit fitting}
\label{sec:fitting}

Similar to our implementation of the baseline \eor model, we also construct our $T_{14}+umb$ model via \texttt{exoplanet} with NUTS sampling and use it to model our synthetic transit signals. This parameterization is motivated by observable transit properties and characterized by the basis $\{P, t_0, r, b, T_{14}\}$. Like the \eor model, the $T_{14}+umb$ model also includes quadratic limb darkening parameters $\{u_1, u_2\}$ as well as fixed values of $\mu_{\rm flux}$ and $\sigma_{\rm flux}$. The priors used here are identical to those used in our \eor model, summarized in Table \ref{tab:priorstab}. Neither $e$ nor $\omega$ is explicitly constrained during the sampling process here, and their values are instead estimated from post-model importance sampling. This parameterization is thus agnostic to orbital eccentricity, except for the implicit assumption of a symmetric transit. This is a reasonable approximation since the acceleration of an eccentric planet during its transit is unlikely to introduce detectable asymmetry given modern photometry \citep{Barnes07}. 

To improve both the sampling convergence and the exploration of complex posterior topologies, we follow \cite{Gilbert2022} to implement umbrella sampling. We separate our NUTS sampler into three windows (i.e. ``umbrellas'') defined within the joint $\{r, b\}$ parameter space, which allows us to sample the full posterior parameter space in smaller pieces that are easier to explore. The resulting posteriors can later be stitched together by applying the appropriate umbrella weights. The three umbrella windows that we use correspond to non-grazing and grazing orbits separated by a region that we refer to as the transition umbrella, which partially overlaps with the other two (see \citealt{Gilbert2022} for full description). In our implementation, we apply the three umbrella models in series but emphasize that this task can easily be parallelized to reduce the apparent wall-clock run-time. In the Appendix, we also discuss a potential alternative to umbrella sampling, known as dynamic nested sampling (see, e.g., \citealt{Skilling2004}; \citealt{Skilling06}), which achieves roughly comparable results.

\subsection{Importance sampling}
\label{sec:estimation}

To recover $\{e, \omega\}$ samples from the $T_{14}+umb$ modeling approach, we apply post-hoc importance sampling to the combined umbrella model posterior distributions. Importance sampling (see, e.g., \citealt{OhBerger93}; \citealt{Gilks95}; \citealt{MadrasPiccioni99}) allows one to measure the properties of a given parameter’s probability distribution based on samples generated from a different (typically easier to sample) parameter’s distribution. This method was first incorporated into exoplanet characterization models by \cite{Ford05} and \cite{Ford06}, used in combination with MCMC sampling to improve radial velocity model efficiency. Such methods can be useful to correct for observational biases post-hoc or derive the distributions of more complicated distributions outside of the MCMC sampling routine. Importance sampling is closely related to umbrella sampling, and the former can be thought of as a single-window special case of the latter. In our implementation, importance sampling only marginally increase the total run-time of the $T_{14}+umb$ approach by a few seconds. 

We first compute the relative weights of the three umbrella models following \cite{Gilbert2022} and combine our posterior chains into a single set of weighted posterior distributions. Since the umbrella weights effectively reduce the total number of samples, we up-sample the merged posterior distributions via random resampling to generate a total of 10$^5$ samples per parameter for convenience. We then perform importance sampling to weigh how well the measured values of $\{P, r, b, T_{14}\}$ at each sampler step can be described by an independently measured density of the host star. We will refer to this independent stellar density as $\rho_{\rm \star, true}$, with some uncertainty $\sigma_{\rho_{\rm \star, true}}$. To determine the appropriate importance weights, we first calculate the sampler-derived stellar density, $\rho_{\rm \star, samp}$, at each point in the umbrella-weighted posterior. This calculation directly follows from the transit duration equation described by \cite{Winn2010}:

\begin{equation}
\label{eq:rhostar}
\rho_{\rm \star, samp} = \frac{3 \pi}{G P^2}\left(\frac{\left(1+r\right)^2-b^2}{\sin^2{\left(\frac{T_{\rm 14} \pi}{P}\frac{1+e\sin{\omega}}{\sqrt{1-e^2}}\right)}}+b^2\right)^{3/2}.
\end{equation}
We note that Equation \ref{eq:rhostar} explicitly includes $e$ and $\omega$, for which we do not yet have any information. We substitute these parameters with random draws of $\{e, \omega\}$ from uniform priors $e \sim U(0,0.92)$ and $\omega \sim U(-\frac{\pi}{2},\frac{3\pi}{2})$ - recall that the upper limit $e=0.92$ was chosen to circumvent sampling issues at high $e$ in the \eor basis. By deriving $\rho_{\rm \star, samp}$ from measured values of $\{P, r, b, T_{14}\}$ and random uniform values of $\{e, \omega\}$, we ensure that $\rho_{\rm \star, samp}$ reflects a true stellar density as opposed to the pseudo-stellar density parameterization which assumes $e=0$ and was deemed unreliable by \textit{G22}.

We compare the samples of $\rho_{\rm \star, samp}$ against the independently measured $\rho_{\rm \star, true}$ by computing the log-likelihood of each $i^{\rm th}$ sample,

\begin{equation}
\label{eq:log-like}
    \log \mathcal{L}_i = -\frac{1}{2}\Big(\frac{\rho_{\rm \star, samp, i} - \rho_{\rm \star, true}}{\sigma_{\rho_{\rm \star, true}}}\Big)^2,
\end{equation}
assuming a Gaussian likelihood function. We then weight each sample from our umbrella-weighted posterior distributions by
\begin{equation}
\label{eq:weights}
    w_i = \frac{\mathcal{L}_i}{\sum_i \mathcal{L}_i}
\end{equation}
to produce the final, importance-weighted posterior distributions for each parameter. We apply these same weights to the random uniform \{$e$, $\omega$\} samples to derive the final posterior distributions of these two parameters. All analysis in this work regarding the $T_{14}+umb$ model is based on these posterior distributions that have been umbrella-weighted, up-sampled, and importance-weighted. The final posterior distribution of $e$ that we measure using our $T_{14}+umb$ modeling approach can thus be directly compared to the $e$ posterior from the \eor model.

\begin{figure*}[ht]
\centering
\includegraphics[width=0.95\textwidth]{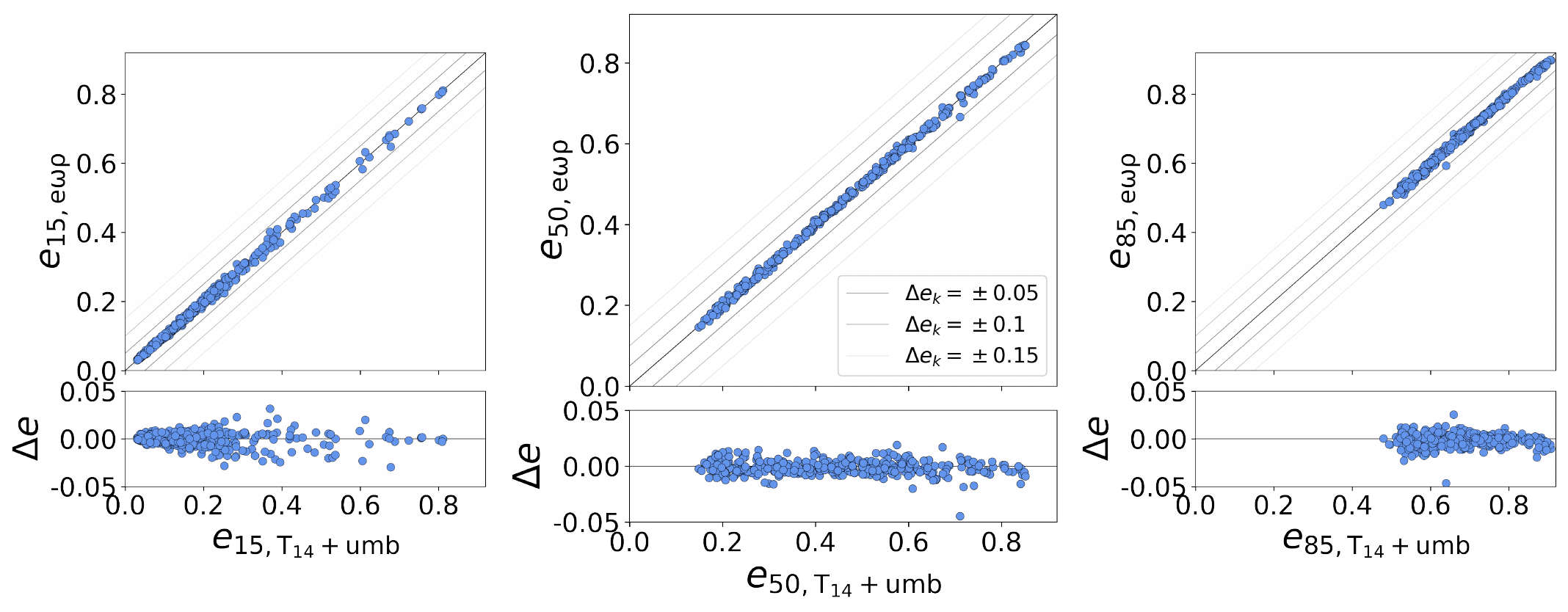}
\caption{Comparison of $e$ values measured from the $T_{14}+umb$ and \eor modeling methods at the 15$^{\rm th}$, 50$^{\rm th}$, and 85$^{\rm th}$ percentiles of their distributions, along with the residuals $\Delta e_{k}$ for each comparison (e.g. $\Delta e_{\rm 50} = e_{50, e\omega\rho} - e_{50, umb}$). We show $\Delta e_k$ = \{0.05, 0.1, 0.15\} in grey, as well as the ideal 1-to-1 line shown in black. These comparisons generally lie close to the 1-to-1 line, implying that the results of the two models are approximately equivalent. We see no trends in the residuals of these comparisons.}
\label{fig:qq-eor}
\end{figure*}

We therefore can use the $T_{14}$ basis $\{P, t_0, b, r, T_{14}\}$ along with an independently constrained $\rho_{\rm true}$ to derive posterior distributions for all parameters represented by the \eor basis $\{P, t_0, b, r, e, \omega, \rho_\star\}$. With the $T_{14}$ basis, we have the advantage of avoiding introducing significant stellar constraints (i.e. $\rho_\star$) until \textit{after} the transit has already been fully modeled. Thus, our $T_{14}+umb$ model only needs to be run once while the \eor model would have to be re-run for each updated measurement of stellar density. The post-hoc importance sampling step can easily be re-run for an updated $\rho_{\rm \star, true}$ value (or different priors on $e$ or $\omega$) within only a few seconds, making our $T_{14}+umb$ modeling approach essentially future-proof. In the era of \textit{Gaia} and high-precision stellar characterization, such future-proofing will become increasingly valuable.

\section{Results}
\label{sec:results}

\subsection{Both methods return equivalent eccentricity constraints}
\label{sec:posteriors}

We fit 375 injected transit signals from our grid of injection-recovery tests using both the \eor baseline model and our $T_{14}+umb$ modeling approach. We measure all transit parameters using both modeling approaches, including $e$ and $\omega$. The posterior distributions of $e$, $\omega$, and $b$ serve as our primary points of comparison between the baseline model and our alternative modeling approach. Here, we specifically focus our analysis on $e$, since $b$ (and its relationship with $r$) was already covered in \textit{G22} and $\omega$ is often a nuisance parameter in photometric modeling. We use posterior comparisons of $\omega$ and $b$ for secondary analysis when necessary.

We perform a quantile-quantile comparison of the posterior values $e_{k}$ at the $k =$ 15$^{\rm th}$, 50$^{\rm th}$, and 85$^{\rm th}$ percentiles of the $e_{e\omega\rho}$ and $e_{T_{14}+umb}$ eccentricity distributions. In Figure \ref{fig:qq-eor}, we present a comparison of $e_k$ from both modeling methods at each of the key percentiles for all injection-recovery tests. We see that all tests at each percentile are close to the 1-to-1 line (black), demonstrating that the two modeling methods produce nearly equivalent posterior results for $e$.

We compute the difference $\Delta e_{k}$ (e.g. $\Delta e_{\rm 50} = e_{50, e\omega\rho} - e_{50, umb}$) and use this as a measure of similarity between the two model results. To estimate the significance of $\Delta e_k$ for each posterior comparison, we assume a standard eccentricity uncertainty of $\sigma_e = 0.05$, informed by the typical uncertainty on $e$ measured among all known planets ($\sigma_{\rm median}(e) \approx 0.05$; \citealt{nea}\footnote{NASA Exoplanet Archive data retrieved on 2023 February 23}). For injection-recovery tests where $|\Delta e_k| \lesssim 0.05$ at the 15$^{\rm th}$, 50$^{\rm th}$, and 85$^{\rm th}$ percentiles of eccentricity, we assert that the \eor and $T_{14}+umb$ methods produce equivalent results. Among multiple iterations of our suite of injection-recovery tests, we did not identify any tests which consistently produced posterior measurements for $e$ that differed by $|\Delta e_k| \lesssim 0.05$ (see Figure \ref{fig:qq-eor}). This suggests that our approach is an excellent alternative to the \eor method, since the two methods should converge on identical results (as opposed to $\sim$68$\%$ identical).

\begin{figure*}[ht]
\centering
\includegraphics[width=0.9\textwidth]{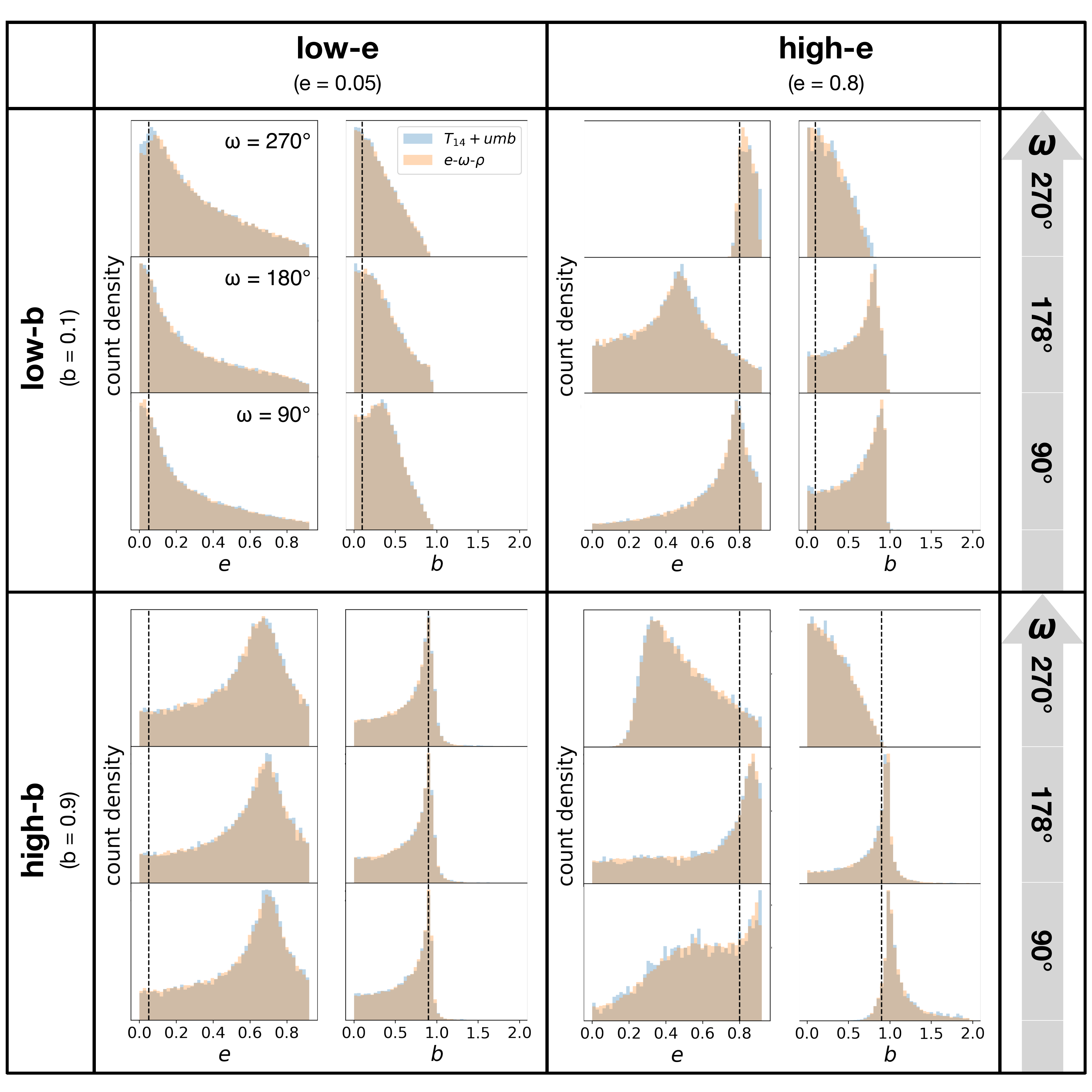}
\caption{Qualitative diagram showing the relative accuracy of measuring $e$ and $b$ from transit photometry in four distinct quadrants of $e-b$ parameter space, at three different $\omega$ values. The four scenarios shown are (Top Left) low-$e$ and low-$b$, (Bottom Left) low-$e$ and high-$b$, (Top Right) high-$e$ and low-$b$, and (Bottom Right) high-$e$ and high-$b$. Across all areas of $e-b$ parameter space, the $T_{14}+umb$ (blue) and \eor (orange) modeling methods perform equivalently.}
\label{fig:punnet}
\end{figure*}

We also consider how $\Delta e_k$ differs as a function of both the lightcurve SNR and the injected transit duration $T_{14}$. Specifically, we consider the ratio between $T_{14}$ and the expected duration of the same planet on a circular, centrally transiting orbit (the reference duration, $T_{\rm 14, ref}$): $T_{\rm 14} / T_{\rm 14, ref}$. This duration ratio is a more concise metric to interpret the effects of $e$, $\omega$, and $b$ on the duration of a transit. While we observe no trend in $\Delta e_k$ with respect to SNR, we do note a marginal trend in $\Delta e_k$ as a function of $T_{\rm 14} / T_{\rm 14, ref}$ across our sample. We find that the $T_{14}+umb$ model estimates slightly higher $e$ values than the \eor model at short transit durations and vice-versa at long transit durations, but the deviations that contribute to this trend are sub-significant. We ultimately conclude that the two modeling methods produce equivalent eccentricity measurements (within a reasonable uncertainty) for virtually all tenable combinations of $\{e, \omega, b, \textrm{SNR}\}$.

\subsection{Both methods return accurate results}
\label{sec:trends-accuracy}

\begin{figure*}[ht]
\centering
\includegraphics[width=0.95\textwidth]{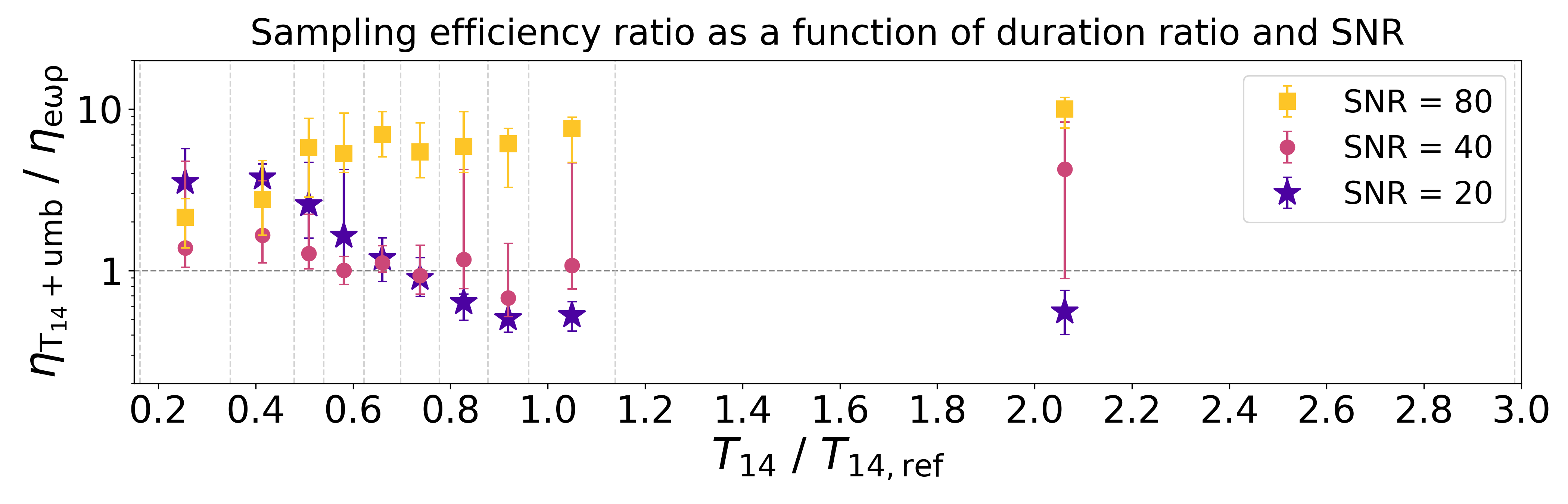}
\caption{Ratio of sampling efficiencies $\eta_{T_{14}+umb}/\eta_{e\omega\rho}$ as a function of duration ratio $T_{\rm 14} / T_{\rm 14, ref}$ and SNR. We bin the data across every 10$^{\rm th}$ percentile of the duration ratio distribution, showing a single point per bin per SNR (bins are separated by vertical grey lines). Each point shows the 15$^{\rm th}$, 50$^{\rm th}$, and 85$^{\rm th}$ percentiles of a given bin. The increased efficiency of the $T_{14}+umb$ method relative to the \eor method depends on the SNR of the modeled lightcurve. At higher SNR the $T_{14}+umb$ method is significantly more efficient, but at moderate-to-low SNR the two methods have more similar efficiencies. At shorter transit durations, the $T_{14}+umb$ method is always more efficient, but this behavior changes around $T_{\rm 14} / T_{\rm 14, ref}$ $\approx 0.8$. The large spread in some uncertainties reflects the heterogeneity of our injected lightcurve parameters.}
\label{fig:eff-eor}
\end{figure*}

We have demonstrated that our alternative transit modeling approach produces equivalently accurate results relative to our baseline model, but we have not yet considered if these models yield the correct results (relative to the injected parameters). It is known in the field of exoplanet characterization that photometric eccentricity constraints (and $\omega$ constraints) tend to have large uncertainties for individual planets (see, e.g. \citealt{VE19}). Here, we qualitatively assess these uncertainties across our set of injection-recovery tests.

Since our sample is not representative of the observed planet population, we describe the observed trends among our $e$ measurements according to different quadrants of $e-b$ parameter space. We split up our tests into four broad scenarios based on their injected transit properties: (1) low $e$ and low $b$, (2) low $e$ and high $b$, (3) high $e$ and low $b$, (4) high $e$ and high $b$. We show demonstrative examples of of these four scenarios in Figure \ref{fig:punnet} with several $\omega$ values, all at SNR $= 20$. In all four quadrants, the posterior distributions of $e$ and $b$ are broad, non-Gaussian, and display a range of outcomes, but we describe the general trends that we observe below. We also offer some additional discussion regarding how $\omega$ can affect these posterior constraints. We limit our discussion to only the posterior distributions of the $T_{14}+umb$ modeling approach since the two approaches produce nearly equivalent results.

In scenario 1 (low $e$ and low $b$), transit models accurately measure low values for both $e$ and $b$ with little posterior mass at higher values (Figure \ref{fig:punnet}, top left), regardless of $\omega$. In scenario 2 (low $e$ and high $b$), models tends to significantly overestimate $e$ but produce more accurate measurements of $b$ (Figure \ref{fig:punnet}, bottom left), regardless of $\omega$. The opposite is true in scenario 3 (high $e$ and low $b$) where $b$ tends to be overestimated while $e$ is measured more accurately (Figure \ref{fig:punnet}, top right), except near apastron where both are measured fairly accurately. In scenario 4 (high $e$ and high $b$), transit models tend to accurately measure high values for both parameters with little posterior mass at lower values (Figure \ref{fig:punnet}, bottom right), except near apastron where neither is measured well. We avoid providing a quantitative description of these observed trends because the non-Gaussian posterior distributions are not well-represented by simple summary statistics.

When $e$ is high (e.g. scenarios 3 and 4), the value of $\omega_{\rm true}$ can significantly impact the posterior constraints on $e$ and $b$ due to the degenerate influence that these parameters can have on the observed transit duration, particularly near apastron. On the other hand, we do not observe any noteworthy trends in model accuracy as a function of SNR. For a typical \textit{Kepler} planet which has low $e$, non-grazing $b$, and $\omega$ closer to periastron, we would generally expect to measure $e$ and $b$ posterior distributions that are somewhat consistent with the true underlying orbital geometry of the planet based on the trends that we observe in Figure \ref{fig:punnet}. In Appendix \ref{sec:dynesty}, we briefly explore whether using a different sampler (i.e. dynamic nested sampling via \texttt{dynesty}; \citealt{dynesty}) might yield even more accurate posterior constraints, but our findings there are inconclusive.

\subsection{Our $T_{14}+umb$ method is more efficient than the $\lowercase{e}$--$\omega$--$\rho$ method}
\label{sec:efficiency}

We have shown that the $T_{14}+umb$ basis can be used as an alternative to the \eor basis, achieving equivalent results while also reducing the number of parameters by two. This parameter reduction should increase the efficiency of the $T_{14}+umb$ model, but this approach also requires three separate sampling runs -- one for each of the three umbrellas. To evaluate the overall model efficiencies, we compared the number of effective samples per second ($\eta$) achieved by each method for all injection-recovery tests. 

For the \eor method, we measure the number of effective samples from the $r$ posterior distribution for each test using Geyer's initial monotone sequence criterion via \texttt{arviz} (\citealt{Geyer92}; \citealt{Gelman13}; \citealt{arviz}). We select $r$ because it is a common output between our models and is less affected by complicated parameter degeneracies. We then divide $N_{\rm eff}$ by the total run-time for this model to achieve the \eor sampling efficiency: $\eta_{e\omega\rho}$. For the $T_{14}+umb$ method, we average $N_{\rm eff}$ of the $r$ posteriors from each umbrella model, weighted by their respective umbrella weights. We divide this weighted average by the sum of the run-times for the three umbrella models (e.g. the CPU run-time) to achieve the overall $T_{14}+umb$ sampling efficiency: $\eta_{T_{14}+umb}$. 

We calculate the ratio of these two efficiencies for all injection-recovery tests and find that $\eta_{T_{14}+umb}/\eta_{e\omega\rho} > 1$ for $\sim$73$\%$ of tests, suggesting that the $T_{14}+umb$ approach is generally more efficient across our set of injected planet parameters. The median value of $\eta_{T_{14}+umb}/\eta_{e\omega\rho}$ across our sample is 2.0, implying that the $T_{14}+umb$ approach is typically 2$\times$ more efficient than the \eor method, although the range of this efficiency ratio is broad. When we consider $\eta_{T_{14}+umb}/\eta_{e\omega\rho}$ as a function of SNR, however, we measure a median efficiency increase of 5.7$\times$ at SNR = 80, 1.2$\times$ at SNR = 40, and 1.1$\times$ at SNR = 20 (see Figure \ref{fig:eff-eor}). We also find that the $T_{14}+umb$ method is only more efficient than the \eor method in $\sim$52$\%$ of low-SNR tests. These findings suggest that the $T_{14}+umb$ method tends to be less efficient when the transit signal is weaker.

From Figure \ref{fig:eff-eor}, we also see that the efficiency ratio changes with respect to the duration ratio $T_{\rm 14} / T_{\rm 14, ref}$. For tests with SNR = 20, the median efficiency ratio $\eta_{T_{14}+umb}/\eta_{e\omega\rho}$ decreases significantly as the duration ratio increases, dropping from 2.1$\times$ at $T_{\rm 14} / T_{\rm 14, ref} \leq 0.8$ to 0.6$\times$ at $T_{\rm 14} / T_{\rm 14, ref} > 0.8$. This trend is likely due to differences in how the two methods explore the high-$b$ grazing regime. As the duration ratio approaches unity or higher, high $b$ values are significantly less likely, but the $T_{14}+umb$ approach continues to carefully explore the high-$b$ regime via three umbrella models even when it is not necessary. On the other hand, injection-recovery tests with higher $b$ values (and generally shorter transit durations) are more efficiently sampled by the $T_{14}+umb$ approach. This behavior is consistent with what we would expect, given that umbrella sampling is specifically intended to ensure accurate measurements of the high-$b$ parameter space.

Our set of injected transit properties, however, is not completely representative of observed planet demographics. To make a more representative comparison, we estimate the efficiency ratio for a typical \textit{Kepler} planet based on both SNR and duration ratio $T_{\rm 14} / T_{\rm 14, ref}$. We use the latter metric because it reflects the combined effects of $e$, $b$, and $\omega$ in a single variable. For a typical confirmed \textit{Kepler} planet with SNR $\approx$ $20-40$ and $T_{\rm 14} / T_{\rm 14, ref} \approx 0.6-1.1$, we estimate an efficiency ratio of $\sim$0.9$\times$. Based on these findings, we assert that the two methods generally have similar sampling efficiencies for real planetary transit signals, with the $T_{14}+umb$ approach excelling for signals with higher SNR or lower duration ratio.

The efficiency increase from the $T_{14}+umb$ approach is more significant when we consider wall-clock time rather than CPU time. Since the three umbrella models can be run in parallel, we can reduce the apparent run-time of the $T_{14}+umb$ approach by up to a factor of a few. In this parallelized case, the apparent sampling efficiency of the $T_{14}+umb$ method is $\sim$1.2$\times$ faster than the \eor method for a typical \textit{Kepler} planet. As another added benefit, the posteriors of the $T_{14}+umb$ approach can be importance sampled for updated values of $\rho_{\star}$ (as they become available) \textit{without} re-running the NUTS sampling process (see \S\ref{sec:estimation}), which is a major advantage in the long-term efficiency of the $T_{14}+umb$ parameterization.

\section{Conclusions}
\label{sec:conclusion}

In this work, we presented an updated photo-eccentric transit modeling method using a duration-based parameterization $\{P, t_0, r, b, T_{14}\}$ (with umbrella sampling) and post-hoc importance sampling which efficiently achieves accurate constraints on $e$, $\omega$, and $b$. Through a suite of synthetic injection-and-recovery tests, we demonstrated that our approach produces equivalent eccentricity constraints relative to the more common eccentricity-explicit transit model parameterization $\{P, t_0, r, b, e, \omega, \rho\}$. We find that our modeling method generally has a higher sampling efficiency than the \eor method when the true $e$ or $b$ value is high or a similar efficiency otherwise. Our approach can also be parallelized to increase its relative sampling efficiency several-fold more.

A key advantage of our modeling method is that post-hoc importance sampling allows us to successfully derive accurate $e$ and $\omega$ posterior distributions (relative to the \eor method) without including $e$, $\omega$, or $\rho$ as explicit model parameters. Our importance sampling routine is fast and flexible enough to easily incorporate an updated prior on $e$ and/or $\omega$, which is critical for hierarchical modeling approaches at the population level. Our method also allows us to update parameter posterior distributions according to updated values of $\rho_\star$ (e.g. from new \textit{Gaia} data releases) without any loss of generality. In the modern era of high-precision stellar characterization, this sort of ``future-proofing'' will be invaluable as the number of transit candidates around well-characterized stars continues to grow.

\begin{acknowledgments}
We are grateful to Dan Foreman-Mackey for helpful conversations about this work. This study made use of computational resources provided by the University of California, Los Angeles and the California Planet Search.

G.J.G., M.G.M., and E.A.P. acknowledge support from NASA Astrophysics Data Analysis Program (ADAP) grant (80NSSC20K0457). E.A.P. acknowledges support from the Alfred P. Sloan Foundation. M.G.M acknowledges support from the UCLA Cota-Robles Graduate Fellowship. 

\end{acknowledgments}

\vspace{5mm}
\facilities{Kepler}

\software{\texttt{astropy} \citep{astropy:2018},
          \texttt{exoplanet} \citep{exoplanet:2021},
          \texttt{numpy} \citep{numpy:2020}, 
          \texttt{PyMC} \citep{pymc3:2016},
          \texttt{scipy} \citep{scipy:2020},
          \texttt{batman} \citep{batman},
          \texttt{arviz} \citep{arviz},
          \texttt{dynesty} \citep{dynesty}
          }

\bibliography{photoecc}
\bibliographystyle{aasjournal}

\appendix

\section{Sampling methods}
\label{sec:mathmeth}

Modeling transit photometry requires efficient exploration of the joint posterior parameter space for some number of transit parameters. A higher number of parameters typically increases the complexity of the posterior space and decreases modeling efficiency. Various tools have been developed for sampling from these complicated posteriors, which we employ and compare throughout this work.

In this section, we briefly review several sampling techniques. This review is not intended to be exhaustive, but rather serves as a jumping-off point for readers who may be unfamiliar with one or more methods explored in this work.

\subsection{Importance sampling}
\label{subsec:math_importance}

Importance sampling (see, e.g., \citealt{OhBerger93}; \citealt{Gilks95}; \citealt{MadrasPiccioni99}) allows one to measure the properties of given parameter’s probability distribution based on samples generated from a different (typically easier to sample) parameter’s distribution. This method was first incorporated into exoplanet characterization models by \cite{Ford05} and \cite{Ford06}, used in combination with MCMC sampling to improve radial velocity model efficiency. Such methods can be useful to correct for observational biases post-hoc or derive the distributions of more complicated distributions outside of the MCMC sampling routine. Importance sampling is closely related to umbrella sampling (see \S\ref{subsec:math_umbrella}), and the former can be thought of as a special case of the latter.

\subsection{Umbrella sampling}
\label{subsec:math_umbrella}

A critical challenge for any sampling problem is knowing when (and if) the posterior space has been fully explored. Even the most sophisticated sampling algorithms may fail to find isolated modes or explore the long tails of distributions. Moreover, convergence tests may offer no hint that portions of parameter space have been missed. In order to ensure proper sampling, one may adopt umbrella sampling \citep{TorrieValleau1977, Kastner2011}, which manually forces the sampler to consider all parts of the posterior topology.

The core idea behind umbrella sampling is straightforward: rather than sampling from a pathological posterior using a single chain (or set of live points), we break the problem into smaller more manageable pieces (``windows,'' in the standard nomenclature), sample from the sub-distributions independently, and then recombine the sub-samples into a single joint posterior distribution after the fact. As long as all sub-distributions are adequately sampled, umbrella sampling will return results that are at least as good as those obtained through standard methods, and often better. Umbrella sampling does not replace other sampling methods, but rather works in tandem with them as a meta-strategy for guiding the sampling problem. Umbrella sampling was introduced into the astrophysics literature by \citet{Matthews2018} and adapted to the problem of exoplanet transits by \citet{Gilbert2022}.

\section{Sampler comparison: NUTS vs. Nested Sampling}
\label{sec:dynesty}

In the previous sections, we demonstrated that the baseline \eor model and our alternative $T_{14}+umb$ approach yield equivalent results when posterior samples are obtained using MCMC methods. Unfortunately, we also saw that posterior inferences of eccentricity can be significantly over- or under-estimated relative to their true values. Here, we explore whether using a different sampling technique -- dynamic nested sampling (\citealt{Skilling2004}; \citealt{Skilling06}) -- can yield more accurate results and/or serve as a potential alternative to NUTS sampling with umbrella sampling.

We implement the duration-based parameterization using the \texttt{dynesty} framework for dynamic nested sampling (\citealt{dynesty}), which does not necessitate the use of umbrella sampling because it already accomplishes the same goal of thoroughly exploring complicated posterior topologies. To model the transit shape and measure the log-likelihood at each sampler step, we use a modified version of \texttt{batman} which takes $\{P, t_0, r, b, T_{14}\}$ as explicit transit parameters (in contrast to the default set $\{P, t_0, r, b, e, \omega, \rho_\star\}$). As before, we perform post-hoc importance sampling to obtain $\{e, \omega, \rho_\star\}$ samples. We apply this alternative modeling method, $T_{14}+dyn$, to all 375 injection-recovery tests in an identical manner as the previous models. 

\subsection{Posterior comparison}
\label{sec:dynesty-posterior}

\begin{figure*}[ht]
\centering
\includegraphics[width=0.95\textwidth]{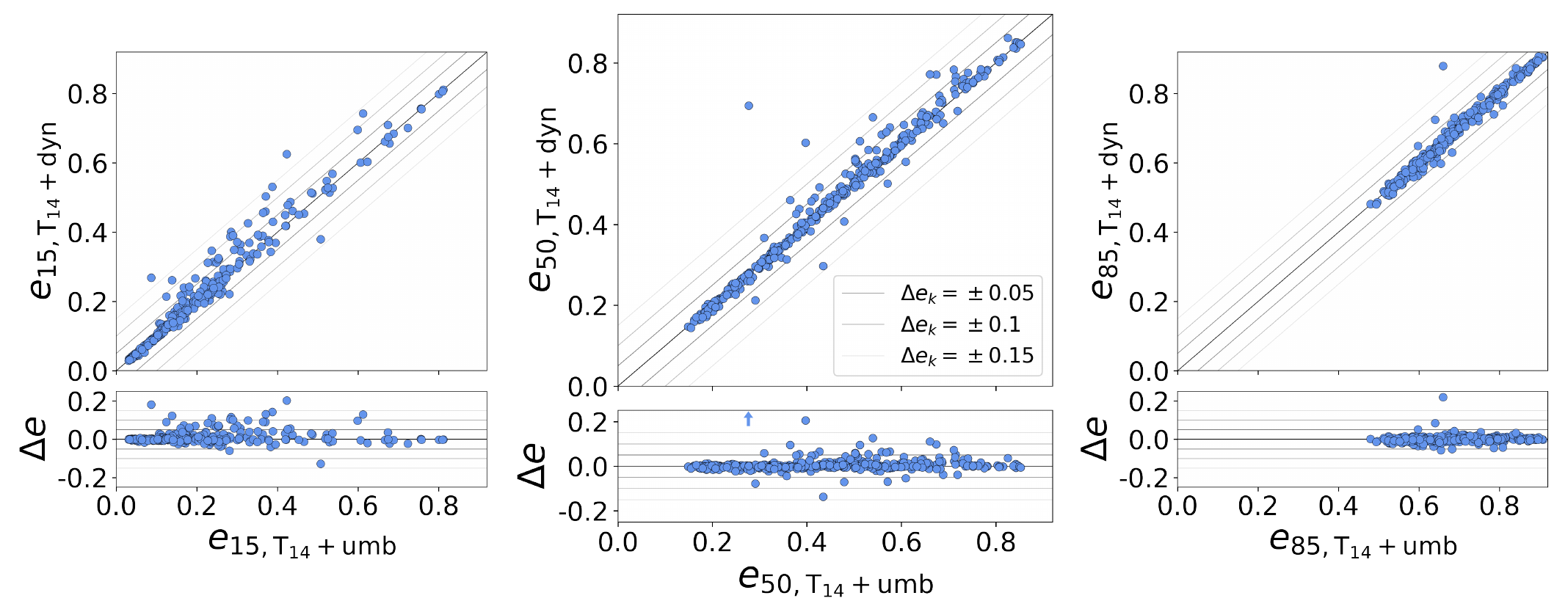}
\caption{Comparison of $e$ values measured from the $T_{14}+umb$ and $T_{14}+dyn$ modeling methods at the 15$^{\rm th}$, 50$^{\rm th}$, and 85$^{\rm th}$ percentiles of their distributions, along with the residuals $\Delta e_{k}$ for each comparison (e.g. $\Delta e_{\rm 50} = e_{50, dyn} - e_{50, umb}$). One outlier residual lies beyond the bounds of our residual plots, as indicated by the arrow pointing towards the outlier at $\Delta e_{50} = 0.417$. We show the $\Delta e_k$ = \{0.05, 0.1, 0.15\} in grey, as well as the ideal 1-to-1 line shown in black. These comparisons generally lie close to the 1-to-1 line, implying that the results of the two models are approximately equivalent. However, we do observe 42 tests where the $T_{14}+umb$ and $T_{14}+dyn$ model results are discrepant by more than $|\Delta e_k| \lesssim 0.05$. In these instances, we find that the $T_{14}+dyn$ model tends to overestimate $e$ relative to the $T_{14}+umb$ model, as seen in the residuals and discussed in \S\ref{sec:dynesty-posterior}.}
\label{fig:qq-dyn}
\end{figure*}

For each injection-recovery test, we measure the values $e_{k, T_{14}+dyn}$ from the $T_{14}+dyn$ eccentricity posterior at the $k =$ 15$^{\rm th}$, 50$^{\rm th}$, and 85$^{\rm th}$ percentiles of the distribution and compare to the results of the $T_{14}+umb$ method like in \S\ref{sec:posteriors} (Figure \ref{fig:qq-dyn}). We find that the $T_{14}+umb$ and $T_{14}+dyn$ methods yield eccentricity results that are broadly in agreement. However, there appears to be more differences between samplers ($T_{14}+umb$ versus $T_{14}+dyn$) than between parameterizations ($T_{14}+umb$ versus \eor). The comparison between parameterizations yielded no test results that consistently differed by $|\Delta e_{k}| \geq 0.05$, but the comparison between samplers yields 42 of such discrepancies. Among these, there are three tests that differ by $|\Delta e_{k}| \geq 0.15$ and yield entirely different posterior topologies for $e$.

The discrepant measurements of $\Delta e_{k}$ are most common at the $k =$ 15$^{\rm th}$ percentile, implying that the two sampling methods differ most at sampling the low-$e$ tail of the eccentricity distribution. We observe that the $T_{14}+dyn$ method produces $e$ posterior distributions with much less posterior weight in the low-$e$ tail as compared to the results of the $T_{14}+umb$ approach. We also see a similar divergence of the two methods in the upper tail of the $b$ posterior distributions. This is consistent with our additional observation that the majority of the discrepancies occur in tests with shorter duration ratios ($T_{\rm 14} / T_{\rm 14, ref} \lesssim 0.5$). Most discrepancies also occur at higher SNR levels, counter to expectations. Together, these criteria for discrepant results only match with $\sim$1$\%$ of observed \textit{Kepler} transit signals, implying that real systems are highly unlikely to fall into this subset.

\subsection{Accuracy}
\label{sec:dynesty-accuracy}

We compare the true underlying eccentricity of each injection-recovery test with the measured posterior distribution of $e$ from the $T_{14}+dyn$ modeling method. Overall, we find that the qualitative trends in $e$ and $b$ measured via the $T_{14}+dyn$ method are roughly equivalent to those measured from the $T_{14}+umb$ method (see \S\ref{sec:trends-accuracy}). We do, however, find a significant difference between the accuracies of the two modeling methods among the three most discrepant injection-recovery tests, where $|\Delta e_{50}| \geq 0.15$. For these discrepant tests, seen as outliers in Figure \ref{fig:qq-dyn}, the $T_{14}+dyn$ method achieves more accurate posterior constraints on both $e$ and $b$. This may suggest that differences between samplers can, in some cases, lead to significant differences in the accuracy of modeled parameters. While all three of these tests have $e_{\rm true} = 0.8$, we unfortunately do not find any discernible rules by which to distinguish when sampler differences will lead to substantial differences in the accuracy of posterior results.

\subsection{Efficiency}
\label{sec:dynesty-efficiency}

\begin{figure*}[ht]
\centering
\includegraphics[width=0.95\textwidth]{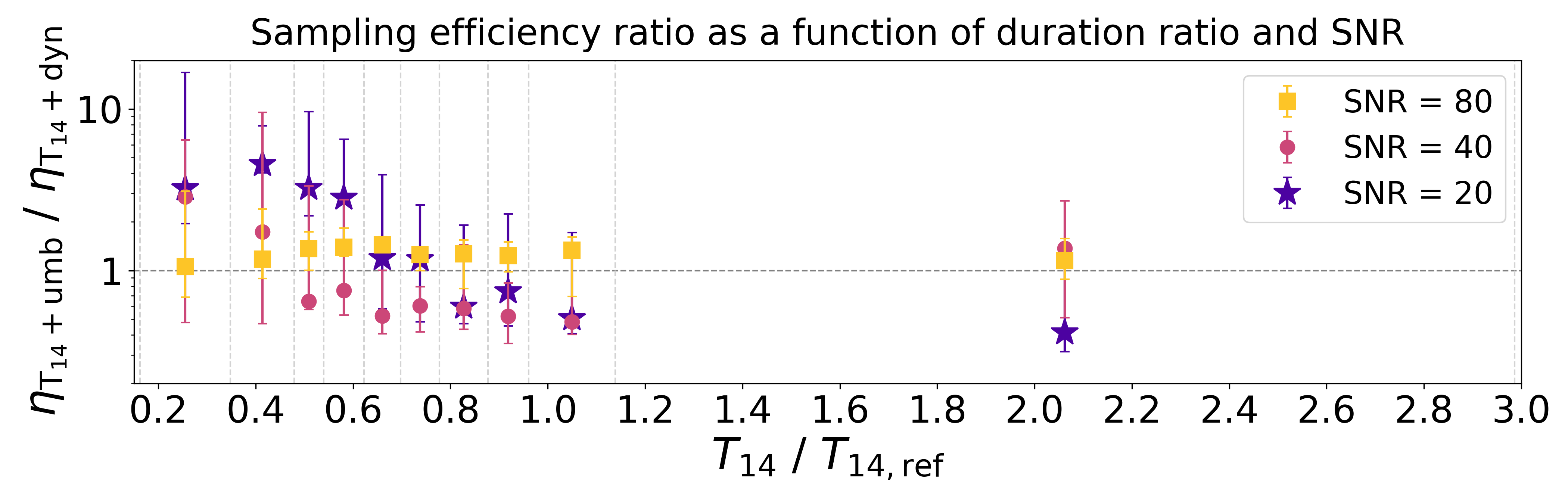}
\caption{Ratio of sampling efficiencies $\eta_{T_{14}+umb}/\eta_{T_{14}+dyn}$ as a function of duration ratio $T_{\rm 14} / T_{\rm 14, ref}$ and SNR. We bin the data across every 10$^{\rm th}$ percentile of the duration ratio distribution, showing a single point per bin per SNR (bins are separated by vertical grey lines). Each point shows the 15$^{\rm th}$, 50$^{\rm th}$, and 85$^{\rm th}$ percentiles of a given bin. The efficiency of the $T_{14}+umb$ method relative to the $T_{14}+dyn$ method depends partially on the SNR of the modeled lightcurve. At shorter transit durations, the $T_{14}+umb$ method is typically more efficient, but the opposite is true at longer transit durations. The large spread in some uncertainties reflects the heterogeneity of our injected lightcurve parameters.}
\label{fig:eff-dyn}
\end{figure*}

We also compare these two modeling approaches according their sampling efficiencies. We calculate the efficiency $\eta_{T_{14}+dyn}$ of the $T_{14}+dyn$ approach for each injection-recovery test, based on the number of effective samples measured via the \citealt{Kish65} approach using \texttt{dynesty}. Similar to \S\ref{sec:efficiency}, we compute the efficiency ratio between the $T_{14}+dyn$ model and our $T_{14}+umb$ approach ($\eta_{T_{14}+umb} / \eta_{T_{14}+dyn}$) and show these results in Figure \ref{fig:eff-dyn}. The distribution of efficiency ratios among our sample is broad but suggests that the two methods generally have similar sampling efficiencies, with a median efficiency ratio of $\eta_{T_{14}+umb} / \eta_{T_{14}+dyn} \approx 1.1$. At lower duration ratios, the $T_{14}+umb$ approach is $\sim$1.4$\times$ more efficient, which is to be expected since this part of parameter space includes higher $b$ values -- the specialty of umbrella sampling as implemented by \cite{Gilbert2022}.

For a typical \textit{Kepler} planet, however, we estimate that the $T_{14}+dyn$ method is $\sim$1.6$\times$ faster than the $T_{14}+umb$ approach. This observation, along with an occasional improvement in accuracy, leans in favor of dynamic nested sampling compared to NUTS sampling $+$ umbrella sampling for our tests, but there are many other compounding factors that are beyond the scope of our experiment. Overall, both sampling methods offer their own benefits with neither winning out 100$\%$ of the time, but it is clear that the duration-based parameterization performs well regardless of the underlying sampling method.

\end{document}